\documentclass[preprint,showpacs,showkeys,superscriptaddress]{revtex4}
\usepackage{graphicx}
\usepackage{amsmath, amsthm, amsfonts,amssymb}

\begin{document}

\title{Relationship between degree of efficiency and prediction in stock price changes}

\author{Cheoljun Eom}
\email{shunter@pusan.ac.kr}
\affiliation{Division of Business Administration, Pusan National University, Busan 609-735, Republic of Korea}

\author{Gabjin Oh}
\email{gq478051@postech.ac.kr}
\affiliation{NCSL, Department of Physics, Pohang University of Science and Technology, Pohang, Gyeongbuk, 790-784, Republic of Korea}

\author{Woo-Sung Jung}
\email{wsjung@physics.bu.edu}
\affiliation{Center for Polymer Studies and Department of Physics, Boston University, Boston, MA 02215, USA}

\date{\today}

\begin{abstract}
This study investigates empirically whether the degree of stock market efficiency is related to the prediction power of future price change using the indices of twenty seven stock markets. Efficiency refers to weak-form efficient market hypothesis (EMH) in terms of the information of past price changes. The prediction power corresponds to the hit-rate, which is the rate of the consistency between the direction of actual price change and that of predicted one, calculated by the \emph{nearest neighbor prediction method} (NN method) using the out-of-sample. In this manuscript, the \emph{Hurst exponent} and the \emph{approximate entropy} (ApEn) are used as the quantitative measurements of the degree of efficiency. The relationship between the Hurst exponent, reflecting the various time correlation property, and the ApEn value, reflecting the randomness in the time series, shows negative correlation ($\rho\left(\overline{H},\overline{A}\right)=-53\%$ where $\rho$ represents cross-correlation, $H$ the Hurst exponent, and $A$ ApEn). However, the average prediction power on the direction of future price change has the strongly positive correlation with the Hurst exponent ($\rho\left(\overline{NN},\overline{H}\right)=86\%$ where $NN$ represents the average prediction power calculated by the NN method)), and the negative correlation with the ApEn ($\rho\left(\overline{NN},\overline{H}\right)=-42\%$). Therefore, the market index with less market efficiency has higher prediction power for future price change than one with higher market efficiency when we analyze the market using the past price change pattern. Furthermore, we show that the Hurst exponent, a measurement of the long-term memory property, provides more significant information in terms of prediction of future price changes than the ApEn and the NN method.

\end{abstract}


\maketitle

\section{Introduction}

The efficient market hypothesis (EMH) has significant influence on theory as well as practical business in the financial literature \cite{Fama}. Various evidence concerning market efficiency has been recently discussed. Also the predictability of future price change in the stock market is a very interesting area in the financial field. Research topics on both the degree of efficiency and predictability generally have been known to be intimately related. In a weak-form EMH, a lower degree of efficiency means that the usefulness of past price change is high in terms of the prediction of future price change. It is difficult to predict the future price change relatively with a higher degree of efficiency. However, the relationship between the degree of efficiency and prediction power has not been empirically studied much relatively. In this study, we investigate the relationship using the stock market indices.

The present study needs a method to quantify the degree of efficiency and prediction
power. The concept of the efficiency corresponds to the weak-form EMH concerning whether the information of past price change pattern is useful in terms of the prediction of future
price change. Therefore, we employed the quantified measurement for the degree of efficiency and the prediction method using its property directly based on the degree of similarity of price change pattern in the time series. Also, we use the Hurst exponent to observe the long-term memory \cite{Hurst}, and approximate entropy (ApEn) to observe
the randomness in the time series \cite{Pincus1}. For the quantitative measurement of prediction power, the hit-rate estimated by the nearest-neighbor prediction method (NN, so called time-delay embedding techniques or a local linear predictor) is used \cite{Farmer,Sauer}. The NN method using reconstructed data which is created by the embedding is useful to predict the future price change based on the degree of similarity of the price change pattern.

The research topic concerning the existence of long-term memory property in the financial time series is generally well-known in the financial field as well as in other scientific fields. When a market has a long-term memory property, the market does not reflect the existing information both immediately and sufficiently. Therefore, if financial time series has a long-term memory property, the information of the past price change becomes a valuable one to predict the future price change. The long-term memory property of the return and volatility in the stock markets has been an active research field \cite{Jacobsen,Hiemstra,Willinger,Grau,Gabjin1}. The Hurst exponent, a measurement of the long-term memory property, quantify the degree of market efficiency of the stock market \cite{Cajueiro,Matteo}. However, the empirical evidence on whether the results of degree of efficiency for long-term memory property directly relates to the prediction power of future price changes based on the patterns of past price changes has not been suggested yet. In this study, we empirically investigate the relationship between the values and prediction power of future price change direction based on the Hurst exponent to quantify the degree of efficiency of stock market.

The ApEn is also a measurement to quantify degree of efficiency in the financial time series, which can quantitatively estimate the degree of randomness in the time series. It calculates quantitatively complexity, randomness, and prediction power. Let us consider the similarity of price change pattern in the financial time series. When the frequency of the similarity pattern is high, the randomness of the time series is low and the ApEn also low. However, the ApEn has a high value if the frequency is low.

Previous works which introduce and apply the ApEn measurement to the financial data \cite{Pincus,Kim} and argue that the ApEn value has significant information in terms of measuring degree of efficiency using the foreign exchange data \cite{Gabjin} are announced. According to the previous studies, the correlation between the Hurst exponent and ApEn value is negative \cite{Pincus,Kim,Cheoljun}. However, the evidence on whether the results of degree of efficiency due to randomness directly relates to the prediction power of future price changes based on the patterns of past price changes has not been suggested yet. In this study, we investigate empirically the relationship between the values and prediction
power of future price change based on the ApEn to quantify the degree of
efficiency of stock markets. 

We also study a prediction method which directly uses the similarity pattern of past price change as the prediction of future price. We employ the NN method to predict future price change based on the similarity of price change pattern. According to the results of previous studies, the NN method is useful to predict the financial time series within a short time frame \cite{Rubio,FFR,Soofi}. In this study, we also investigate the relationship between the prediction power estimated by th NN method and the Hurst exponent, the ApEn, respectively. This investigation observes the relationship between prediction power and degree of efficiency in a financial time series. The prediction power that we use in this paper is a hit-rate, which quantify the consistency between the actual price change and the predicted one by the NN method.

According to the established research purpose and previous studies, we originally expected a positive relationship between the Hurst exponent and prediction power. In other words, a market index with a higher Hurst exponent shows averagely a higher prediction power than one with a lower Hurst exponent. Also, we expected a negative relationship between the ApEn value and prediction power; a market index with a lower ApEn value has averagely a higher prediction power than one with a higher ApEn. Through this study, we find the following results. First, the relationship between the average Hurst exponent $(\overline{H_j})$ as the long-term memory property and the average ApEn value $(\overline{A_j})$ as the randomness is negative, $\rho(\overline{H_{j}}, \overline{A_{j}}) = -53\%$ where $\rho$ represents cross-correlation and $j$ each stock market. Second, the average prediction power $(\overline{NN_{j}})$ in terms of the direction of future price changes reveals a positive relationship with the Hurst exponent $(\rho(\overline{NN_{j}}, \overline{H_{j}}) = 86\%)$, and has a negative one with the ApEn $(\rho(\overline{NN_{j}}, \overline{A_{j}}) = -42\%)$. According to the above results, we find that the relationship between the degree of efficiency and the prediction power is negative. In other words, a market index with a lower degree of efficiency has a higher predictability of the future price change in terms of the past price change pattern than one with a higher degree of efficiency. Moreover, we find that the Hurst exponent as the measurement of long-term memory property provides valuable information in terms of the prediction of future price change.

In the next section, we describe the data and methods of the verification process used in this paper, In section \ref{sec:RESULTS}, we present the results obtained according to the established research purpose. Finally, we summarize the findings and conclusions of the study.

\section{Data and methods}

\subsection{Data}

We study the daily indices of 27 stock markets, which obtained from DataStream (\textit{http://www.datastream.net}). The market indices are composed of 13 markets in Asia-Pacific region, 7 in North and South America, and 7 in Europe. They are China (Shenzhen Composite), Hong Kong (Hangseng), India (Bombay SE200), Indonesia (Jakarta SE Composite), Japan (Nikkei225), Korea (KOSPI200), Malaysia (Kuala Lumpur Composite), Philippines (SE Composite), Singapore (Straits Times), Taiwan (SE Weighted), Thailand (Bangkok S.E.T), Australia (ASX), New Zealand (NZX), Argentina (Base General), Brazil (Bovespa), Chile (General), Mexico (IPC), Peru (Lima SE General), Canada (TSX Composite Index), USA (S\&P 500), Austria (ATX), Denmark (Copenhagen KBX), France (CAC 40), Germany (DAX 30), Italy (Milan Comit General), Netherlands (AEX Index), and the UK (FTSE100). The data period is 15 years from January 1992 to December 2006 except Australia (from June 1992), Argentina (from June 2000), and Denmark (from January 1996). The return time series of the market index used in this paper was calculated by the logarithmic change of the price, ${R(t)} = \ln{P(t)}-\ln{P(t-1)}$ where $P(t)$ is the stock price on day $t$.

\subsection{Method of Hurst Exponent}

In the previous studies, researchers utilized various methods such as the Hurst exponent \cite{Granger,Barkoulas,Kilic}, ARFIMA (autoregressive fractional integration moving average) \cite{Granger1} and FIGARCH (fractionally integrated generalized autoregressive conditional heteroscedasticity) \cite{Baillie} to quantitatively observe the long-term memory property in a financial time series. However, the long-term memory property in a financial time series does not depend on the method used. We use the Hurst exponent to quantify the degree of long-term memory property. There are many methods to calculate the Hurst exponent. They are the classical re-scaled range analysis \cite{Hurst}, generalized Hurst exponent method \cite{Matteo}, modified R/S analysis \cite{Lo}, GPH method \cite{Geweke}, and the detrended fluctuation analysis (DFA) \cite{Peng}. According to the previous works, the DFA method is the most efficient method for accurately calculating the Hurst exponent \cite{Weron,Cheoljun1}. Therefore, we select the Hurst exponent calculated by the DFA method as a measurement to quantify the degree of efficiency in a financial time series.

The Hurst exponent calculated by the DFA method can be explained by the following steps. First, after the subtraction of the mean $(\overline{x})$ from the original time series $(x(i))$, one accumulates the series $(y(i))$ as defined by

\begin{gather}
y(i)= \sum_{i=1}^{N} [x(i) - \overline{x}], \tag{1a}
\end{gather}
where $N$ is the number of the time series. Next, the accumulated time series is divided into a window of the same length $n$.  We estimate the trend lines $(y_{n}(i))$ by using the ordinary least square (OLS) in each window. We eliminate trends existing within the window by subtracting from the accumulated time series in each window. This process is applied to every window and the fluctuation magnitude $(F(n))$ is defined as

\begin{gather}
F(n) = \frac{1}{N} \sum_{i=1}^{N}(y(i) - y_{n}(i))^2. \tag{1b}
\end{gather}

The process mentioned above is repeated for every scale $(n, 2n,3n, \ldots, dn)$. In addition, we investigate whether the scaling relationship exists for every scale. That is, the scaling relationship is defined by

\begin{gather}
F(n) \approx c \cdot n^{H}, \tag{1c}
\end{gather}
where $c$ is the constant and $H$ is the Hurst exponent. If $0 \leq H < 0.5 $, the time series has a short-term memory. If $0.5 < H \leq 1 $, it has a long-term memory. We use the Hurst exponent as the measurement to quantify the degree of market efficiency of a stock market in each market. As the Hurst exponent increases, the persistence of similarity patterns in the past price changes is high. In other words, the pattern of past price change is a valuable information when predicting the patterns of future price change. 

In addition, in order to estimate and use the robust Hurst exponent regardless of the time variation, we use the average Hurst exponent,$(\overline{H_{j}} = \frac{1}{T-1} \sum_{t=1}^{T-1} H_t^{j})$, estimated repeatedly until December 2005 $t=T-1$ by estimating the time window with the width of 5 years and shifting 1 year for a whole period until December 2006 $(t=T)$. We omit the data of the year 2006 when calculating the mean Hurst exponent because an out-of-sample of the 1-year prediction period is required in the NN method.

\subsection{Method of Approximate Entropy Measurement}

The ApEn, which measures the degree of randomness in the time series, was also introduced as the measurement of the degree of efficiency of the financial time series \cite{Pincus,Kim,Gabjin}. We use the ApEn as the second measurement of the degree of efficiency in this study. The ApEn is defined as

\begin{gather}
ApEn(m,r)= \Phi^{m}(r)-\Phi^{m+1}(r), \tag{2a}
\end{gather}
where $m$ is the embedding dimension and $r$ is the tolerance to determine the similarity between price change patterns. The $\Phi^{m}(r)$ is given by

\begin{gather}
\Phi^{m} (r) = (N-m+1)^{-1} \displaystyle \sum_{i=1}^{N-m+1} \ln[C_{i}^{m}(r)], \tag{2b} \\
C_{i}^{m} (r) = \frac{\displaystyle B_{i}(r)}{(N-m+1)}, \tag{2c}
\end{gather}

where $B_{i}(r)$ is the number of data pairs within a tolerance of similarity $r$. Also, we calculate the similarity in the time series of each price change pattern $(u(k), k=1,2,\ldots, m)$ by the distance $d[x(i), x(j)]$ between two vectors $x(i), x(j)$ defined as

\begin{gather}
B_{i} \equiv d[x(i), x(j)] \leq r. \tag{2d} \\
d[x(i),x(j)]= \underset{k=1,2,..,m}{max}(|u(i+k-1)-u(j+k-1)|).
\tag{2e}
\end{gather}

Through the above equations, the ApEn compares the relative magnitude between repeated pattern occurrences for the embedding dimensions $m$ and $m+1$. In other words, the ApEn becomes smaller as the frequency of similar price change patterns for the embedding dimension $m$ equals that for the embedding dimension $m+1$. As both frequencies are same, the ApEn value is 0. On the other hand, ApEn is relatively larger as long as the frequency of similar price change patterns for the embedding dimension $m+1$ is small. As the ApEn is small, the frequency of the similar price change pattern is small, the time series data has a low degree of randomness, and the efficiency of financial time series becomes low. In order to calculate the ApEn, we use the embedding dimension, $m=2$, and tolerance of similarity, $r=20\%$ of standard deviation of the time series, similar to previous works \cite{Pincus,Kim,Gabjin}.

We establish the estimated period to calculate the ApEn as the same process for the Hurst exponent. In order to use the ApEn robustly regardless of time variation, we use the average ApEn value, $(\overline{A_{j}} = \frac{1}{T-1} \sum_{t=1}^{T-1} A_t^{j})$, estimated repeatedly until December 2005 $(t=T-1)$ by estimating a period of 5 years and shifting 1 year for the whole period until December 2006 $(t=T)$. The reason we omit the data during 2006 when calculate the mean ApEn is that we need an out-of-sample of the 1-year prediction period not included in the estimated period of the NN prediction method.

\subsection{Method of Nearest Neighbor Prediction}

In this section, we introduce the NN prediction method, which investigates the relationship between degree of efficiency as the Hurst exponent or the ApEn and the predictability of future price change. In order to observe the relationship between the degree of efficiency and predictability, a prediction method that uses a similar pattern of past price changes should be used as we used the information of similar patterns in the price change to calculate the Hurst exponent and ApEn value. For this reason, we used the NN prediction method. Many previous studies have confirmed the usefulness of the NN prediction method \cite{Rubio,FFR,Soofi}, and the process of the NN prediction method can be expressed as follows.

In the first step, we reconstruct the pattern series, $V_n^{m,\tau}$, with the embedding dimension, $m$, and the time delay, $\tau$ from the financial time series, $x_1, x_2, \ldots, x_T$ and defined as

\begin{gather}
V_{n}^{m,\tau} = [x_n, x_{n-\tau}, \ldots, x_{n-(m-1)\tau}]
\tag{3a} \\
n = (m-1)\tau+1,\ldots, T \notag
\end{gather}

The structure of the reconstructed time series can be explained as follows. The length of the reconstructed pattern series $V_{n}^{m,\tau}$ is $m$ in a time series of $N$ length, and this creates the $N-m+1$ data points. In order to predict price change at time $t+1$, we have to find price change patterns similar to the target price change pattern $V_{Target}^{m,\tau}$ at time $t$ among a $N-m+1$ past price pattern series $V_{n}^{m,\tau}$. The criteria to select the similar patterns in the price
change is the distance between both vectors.

\begin{gather}
D=(V_{Target}^{m,\tau}-V_{n}^{m,\tau})^2. \tag{3b}
\end{gather}

When the distance is 0, both price change patterns are completely identical. As the distance increases, they do not have similar patterns of price change. We choose the $K$ number of similar pattern $(V_{n,k}^{m,\tau}, k=1,2,\ldots,K)$, which has the smallest distance, among the whole pattern series. To confirm whether the chosen similar patterns of past price changes are true, we select the chosen similar patterns $(V_{n,k(\ast)}^{m,\tau}, k(\ast)=1,2,\ldots,K^{\ast})$ for the embedding dimension, $m+1$. The future price change $(x_{n+1})$ can be predicted by the next price change $(F(V_{n,k(\ast)} ^{m,\tau}))$ of the chosen patterns of the price change $(K^{\ast})$ and defined as

\begin{gather}
x_{n+1} = F(V_{n,k(\ast)} ^{m,\tau}). \tag{3c}
\end{gather}

We calculate the hit-rate by the degree of consistence as the prediction power by comparing the direction of actual price change with direction of predicted price change. In addition, in order to employ the instinctive method in the estimation of prediction power, we determine the direction by comparing it to the frequency of direction of past price change patterns. That is, the direction of future price $(x_{n+1})$ is determined by the direction having a higher frequency using the number of direction frequency on whether the next day's price change of the chosen past patterns in the Eq. (3c) is in an upward direction or downward direction.

We use the same period described above to calculate the Hurst exponent and the ApEn. In other words, the period to reconstruct is 5 years. The prediction period is 1-year of out-of-sample, not including the reconstruction period. The prediction day is one day. That is, using the first reconstruction period $(t=1)$, we predict the direction (up or down) of price change for the first trading day $(t_p=1)$ in the prediction period (one year). And, we valuate the consistency between actual price change and that estimated by the NN model. 
In order to predict the direction of price change on the second trading day $(t_p=2)$, we use the pattern series by removing the farthest trading day (1 day) and adding the new trading day (1 day). This process iss repeated for the prediction period (1 year) $(t_{p}=1,2,\ldots, T_p)$. After the prediction period closes, the prediction power can be calculated by the frequency ratio $(NN_{t=1}^{j}=FQ_{t}[T_p]^{(\ast)}]/FQ_{t}[T_p]])$ of the number of consistent direction trading $(T_p^{(\ast)})$ with the number of trading day $(T_p)$ for the prediction period. Next, according to the above process after the moving period (1 years) as the process for the Hurst exponent and the ApEn, we calculate the hit-rate $NN_{t=2}^{j}$ of future price change using the second reconstruction period $(t=2)$. This process is repeated until the reconstruction period $t=T-1$ of the last day of the prediction period. We use the average hit-rates $(\overline{NN_{j}}=\frac{1}{T-1} \sum_{t=1}^{T-1} \overline{NN_t^{j}}, t=1,2,\ldots,T)$ in terms of the consistent direction between actual price changes and those estimated by the NN model as the prediction power to observe the relationship with the Hurst exponent and ApEn as the degree of efficiency.

\section{Results}
\label{sec:RESULTS}

We present the observed relationship between the Hurst exponent and the ApEn to measure the degree of efficiency and the relationship between degree of efficiency and prediction power.

\subsection{The Relationship Between Measurements of Degree of Efficiency}

\begin{figure}

\includegraphics[width=1.0\textwidth]{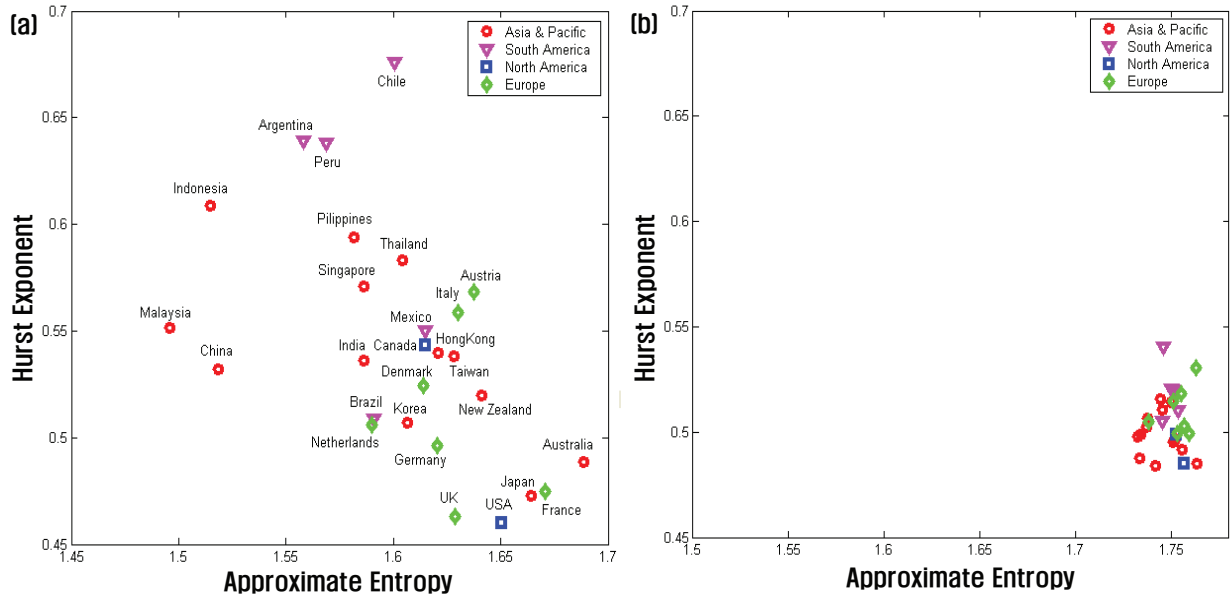}

\caption[0]{The figure shows the relationship between the Hurst exponent as the long-term memory and ApEn as the randomness. (a) and (b) show the results for using a time series of the actual stock market and using a random time series, respectively. The circles (red), triangles (magenta), squares (blue), and diamonds (green) indicate the market indices for the Asia-Pacific region (13 markets), South America (5 markets), North America (2 markets), and Europe (7 markets).}

\end{figure}

In this section, using the stock market indices of 27 markets, we present the investigated results for the relationship between the Hurst exponent and the ApEn value in terms of degree of efficiency. As mentioned above, we expected a negative relationship between the Hurst exponent and ApEn. In other words, from the viewpoint of the Hurst exponent, a lower efficiency degree means a higher Hurst exponent which reveals long-term memory. On the other hand, a lower efficiency degree concerning the ApEn means lower ApEn values, which indicate a lower degree of randomness. Of course, based on both measurements, a lower efficiency degree occurs when similar patterns occurs repeatedly in the past price change. Futhermore, the observed results are negative evidence of the weak-form EMH. The results are presented in Fig. 1.

Fig. 1 shows the average Hurst exponent $(\overline{H_j}=\frac{1}{T-1} \sum_{t=1}^{T-1}H_t^j)$ and the ApEn $(\overline{A_j}=\frac{1}{T-1} \sum_{t=1}^{T-1}A_t^j)$ calculated repeatedly by using the estimated period (5 years) and moving period (1 year) for each stock indices $(j=1,2,\ldots,27)$. In Fig. 1, the x-axis and y-axis indicate the ApEn and Hurst exponent, respectively. Fig. 1(a) and (b) show the results for using a  time series of the actual stock market and using a random time series with the mean and standard deviation of time series of the actual stock indices, respectively. In the case of a random time series, the Hurst exponent and ApEn also have an average value $(\overline{H_{RM(j)}}, \overline{A_{RM(j)}})$ for the whole period. The market index of each country denotes according to a continent. That is, the circles (red), triangles (magenta), squares (blue), and diamonds (green) indicate the Asia-Pacific region (13 countries), South and North America (7 countries), and Europe, respectively.

According to the results, we find that the relationship between the Hurst exponent and the ApEn is negative $[\rho(\overline{H_j}, \overline{A_j})=-53\%]$. However, we could not find these relationships in the random time series. That is, the ApEn in terms of the randomness of the price change patterns is low as the financial time series has a high long-term memory property. Therefore, both methods can be used complementarily in terms of measuring the degree of efficiency in a financial time series. As the market indices locate at the top left, they have either a higher long-term memory or lower randomness. This shows a low degree of efficiency in the time series. We discovered that most stock indices for South America or Asia belong to this area. On the other hand, as the market indices locate at the bottom-right, they have lower long-term memory or higher randomness. That is, this demonstrates a high degree of efficiency. Most stock indices in this position are in Europe (France, the UK, Germany), North America (Canada, the USA), and Asia (Australia, Japan, New Zealand and Korea).

\begin{figure}

\includegraphics[width=1.0\textwidth]{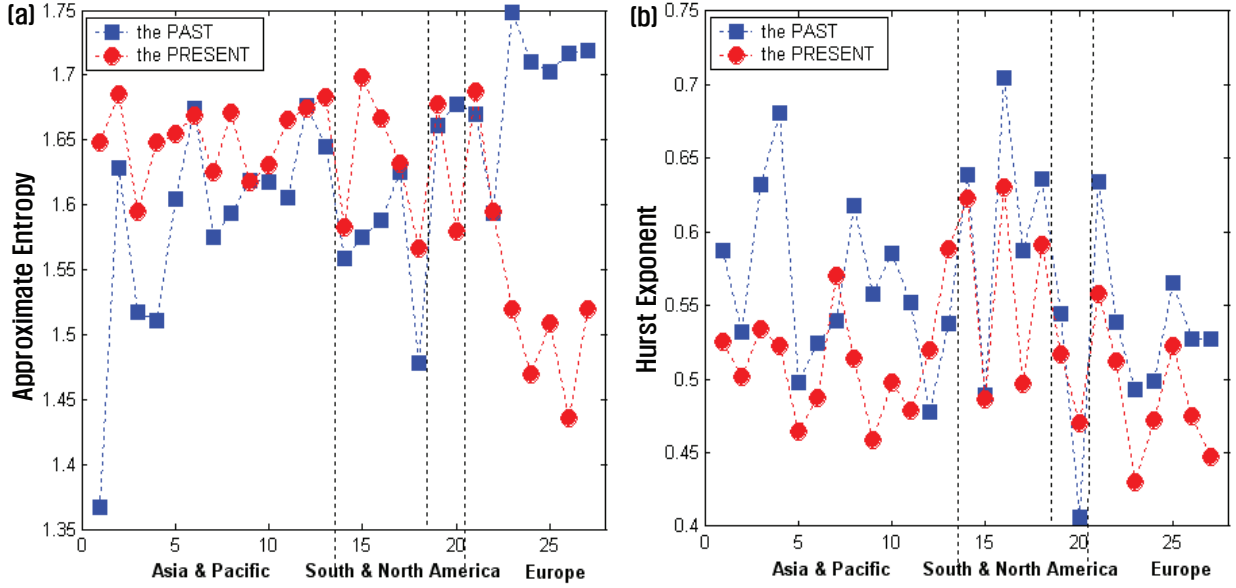}

\caption[0]{The figure shows the Hurst exponent and ApEn calculated at the farthest and nearest past period among various sub-periods used in this analysis. (a) and (b) present the Hurst exponent and ApEn, respectively. The circles (red) and squares (blue) indicate the measures of the farthest and nearest past period, respectively. Futher, the market indices denote by the Asia-Pacific region, South and North America, and Europe, respectively. }

\end{figure}

In addition, we investigated whether the degree of efficiency
increases more in the past period in each market indices. The
results are presented in Fig. 2. Based on the Hurst exponent and
ApEn estimates using each sub-period $(t=1,2,\ldots,T)$ including
2006, we observed measurements at times $t=1$ and $t=T$. The first
sub-period $(t=1)$ for the stock indices is 5 years, from January
1992 to December 1996, except for three countries, Australia
(June, 1992), Argentina (June, 2000), and Denmark (January, 1996).
The last sub-period $(t=T)$ is 5 years, from January 2002 to
December 2006. Figs. 2(a) and (b) show the ApEn $(A_{t=1}^{j},
A_{t=T}^{j})$ and Hurst exponents $(H_{t=1}^{j}, H_{t=T}^{j})$ of
the market indices of 27 countries. In Fig. 2, the first ($t=1$,
the PAST) and last ($t=T$, the PRESENT) sub-periods are denoted by
squares (blue) and circles (red).

By the results, on the whole, we find that the degree of
efficiency in market indices increase nowadays more than in past
periods. In other words, the Hurst exponent of Asia decreased from
0.56 to 0.51, and South-America decreased from 0.61 to 0.57,
respectively. The ApEn for Asia increased from 1.59 to 1.65, for
South-America from 1.56 to 1.63. On the other hand, the Hurst
exponent and ApEn of stock indices for Europe reveal a different
result. That is, the Hurst exponent of past and present approaches
0.5 averagely, but the ApEn decreases averagely from 1.69 to 1.53.
We could not find consistent changes in terms of the degree of
efficiency. Through the results, we confirmed that both methods
can be a measure of the degree of efficiency of a financial time
series. In addition, we can complementarily use both methods as
the standard index of the degree of efficiency because the
relationship between both methods is negative. Although we could
not find consistent results between ApEn measurement and Hurst
exponents for Europe, the present efficiency degree of the market
index for each country increased more than those of past periods.

\subsection{The Relationship Between Degree of Efficiency and Prediction Power}

\begin{figure}

\includegraphics[width=1.0\textwidth]{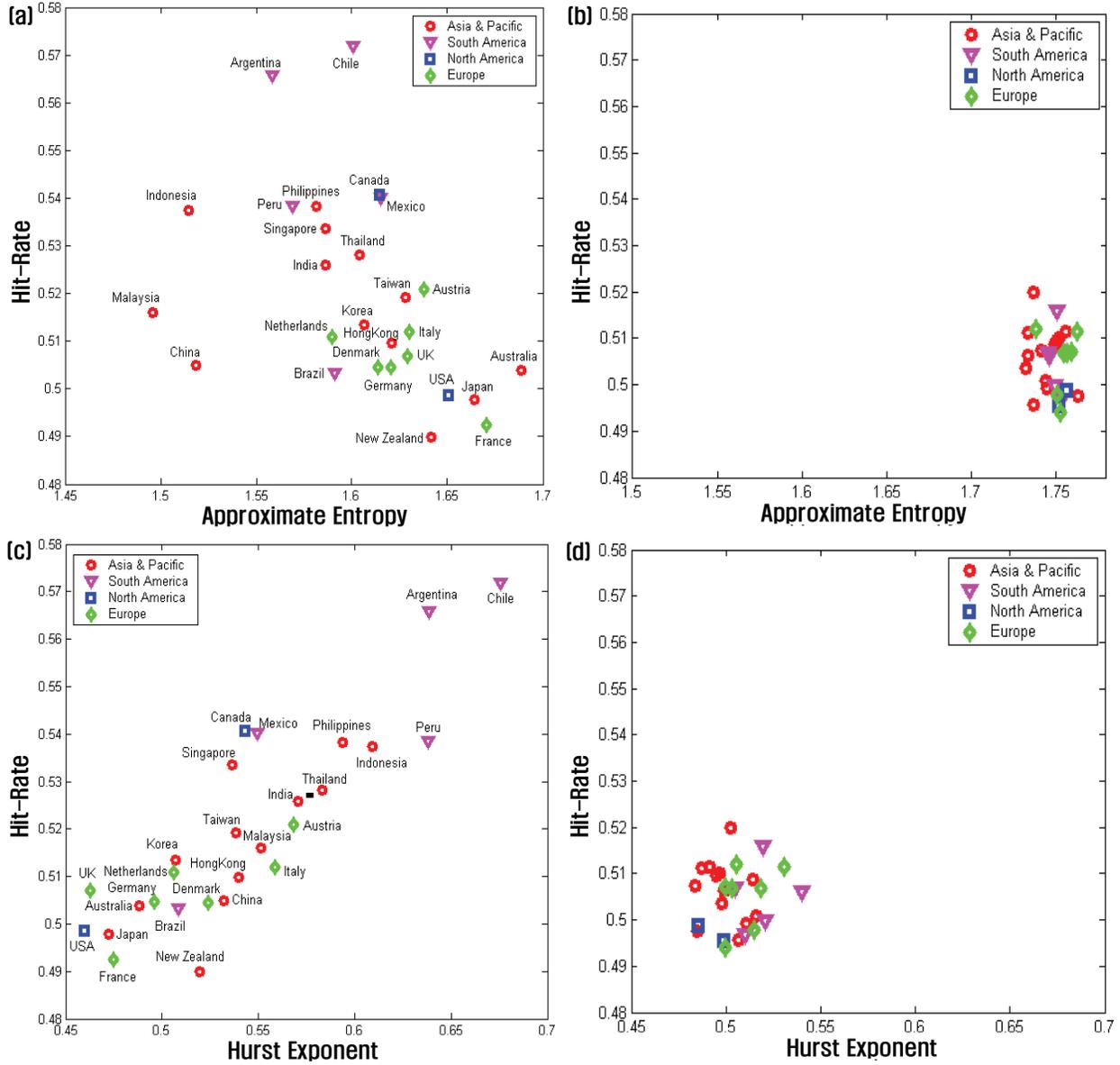}

\caption[0]{The figure contains empirically observed results of the
relationship between degree of efficiency and prediction power.
(a) and (b) show the relationship between prediction power and
ApEn value and (c) and (d) display the relationship between Hurst
exponent and prediction power. We present the results using the
indices of actual stock market in (a) and (c) and the random time
series in (b) and (d). In Fig. 3, the circles (red), triangles
(magenta), squares (blue), and diamonds (green) indicate the
market indices including Asia-Pacific region (13 countries), South
America (5 countries), North America (2 countries), and Europe (7
countries). }

\end{figure}

In this section, we present the results of the empirical
relationship between degree of efficiency and prediction power of
future price changes estimated by the NN prediction method.
Generally, the predictability of a time series having a lower
degree of efficiency is higher than that of a time series having a
higher degree of efficiency. Therefore, the expected relationship
between the Hurst exponent and ApEn values in terms of the degree
of efficiency and prediction power estimated by the NN prediction
method is as follows. We expect a positive relationship between
the Hurst exponent and prediction power of the NN prediction
method. That is, a market index having a higher Hurst exponent
value has a high long-term memory property because there are
continuously similar patterns of past price change. Therefore, the
predictability of future price change using the patterns of the
past price changes when the market index increases. However, we
expect a negative relationship between ApEn and prediction power.
That is, a market index having a lower ApEn has a lower randomness
because the similar patterns of past price changes happen
frequently. Therefore, the predictability of the future price
change using the patterns of the past price changes as the market
index increases. The results are presented in Fig. 3.

In Fig. 3, the prediction power of the NN prediction method is an
average value $(\overline{NN_j}=\frac{1}{T-1} \sum_{t=1}^{T-1}
\overline{NN_{t}^{j}})$ of the hit-rates in terms of the direction
of future price changes in each sub-period. Fig. 3(a) shows the
relationship between average ApEn $\overline{A_j}$ and average
prediction power $\overline{NN_j}$ and Figure 3(c) shows the
relationship between average Hurst exponent $\overline{H_j}$ and
average prediction power $\overline{NN_j}$. Moreover, using the
random time series created with the mean and standard deviation of
the original market index, the relationship between NN prediction
power and ApEn or Hurst exponent is presented in Figs. 3(b) and
3(d), respectively. The market index of each country is denoted
according to continent and color. The Asia-Pacific region, South and North America , and Europe are
denoted by circles(red), triangles(magenta), squares(blue), and
diamonds (green), respectively.

According to the results, we can confirm the relationship between
degree of efficiency and prediction power, generally acknowledged
in the financial literature. That is, in Fig. 3(a), the
relationship between average hit-rates estimated by the NN method
and average ApEn is negative $\rho (\overline{NN_{j}},
\overline{A_{j}} = -42\%)$, while in Fig.3(c), the relationship
between average hit-rates and average Hurst exponent shows a
strong positive relation $\rho (\overline{NN_{j}},
\overline{H_{j}} = 86\%)$. Therefore, stock indices with a lower
degree of efficiency (higher Hurst exponent or lower ApEn) have a
higher prediction power in terms of the direction of future price
change than those with the higher degree of efficiency. Of course,
we could not find Figs. 3(b) and 3(d) using the random time
series. The places which have a lower degree of efficiency and
higher predictability are shown in the top left of Fig. 3(a),
which displays a relationship between ApEn and prediction power
and the top right of Fig. 3(c), which shows a relationship between
the Hurst exponent and prediction power. These market indices are
generally found in South America and Asia. However, places which
have a higher degree of efficiency or lower predictability are
shown in the bottom right of Fig. 3(a) and the bottom left of Fig.
3(b). The market indices displayed here are found in Europe
(France, the UK, Germany \textit{et al.}), North America (the USA), and
Asia (Australia, Japan, New Zealand \textit{et al.}).

According to the above results, we discovered that the degree of
efficiency of a financial time series is significantly related to
the prediction power of future price changes. We also found that
the Hurst exponent and ApEn provide significant information in
terms of the NN prediction method using the similarity of past
price change patterns. In particular, we found that the Hurst
exponent used to measure long-term memory property is closely
correlated with the prediction power of the NN prediction method.

\section{Conclusions}
\label{sec:CONCLUSIONS}

We empirically investigated the relationship between the degree of efficiency and the prediction power using the market index for various countries. According to the viewpoint of the weak-form EMH, a lower degree of efficiency means that the information of the past price changes is useful in terms of the predictability of future price change, and a higher efficiency degree demonstrates that the predicting is difficult relatively. The efficiency in this paper is based on the weak-form EMH which used to show whether the information of past price changes patterns is useful in terms of the prediction of future price changes. 
We employed the measurements such as the Hurst exponent and the ApEn which quantify the degree of efficiency and the prediction method (NN prediction method) by using the measurements. The measurements and the method are based on the similarity of price change pattern in the time series.

We summarize the results as follows. First, we found that the relationship between the Hurst exponent and the ApEn, the measurements of the degree of efficiency, is negative. Therefore, the Hurst exponent and the ApEn are useful to measure the degree of efficiency in the financial time series complementally. Second, we found that the degree of efficiency is positively related to the prediction power. Market indices having a lower degree of efficiency (high Hurst exponent or low ApEn) have a higher prediction power  in terms of the direction of the future price change than those having a higher degree of efficiency (low Hurst exponent or high ApEn). Futhermore, we found that the Hurst exponent has a very strong correlation with the prediction power of NN prediction power.

According to the above results, we \textit{empirically} confirm the relationship between the degree of efficiency and the prediction power, which have acknowledged \textit{qualitatively} in the financial field. We also found that the Hurst exponent and the ApEn provide useful information to the NN prediction method using the similarity of past price change pattern. In addition, our study provides useful information in terms of the consistency of the international funds which have considerable attention in the financial field and the selection of active and passive investment countries as investment strategy, and the investment weights for future works. Also, further works as the complementary process are expected.


\end{document}